\documentclass[11pt]{article}

\newcommand{\be}{\begin{equation}}
\newcommand{\ee}{\end{equation}}
\def\bea{\begin{eqnarray}}
\def\eea{\end{eqnarray}}
\def\der{\partial}

\newcommand{\wh}{\widehat}
\newcommand{\mscr}[1]{\mbox{\scriptsize #1}}

\newcommand{\ft}[2]{{\textstyle\frac{#1}{#2}}}

\newcommand{\Smacro}{{\mathcal S}_{\mscr{macro}}}
\newcommand{\Smicro}{{\mathcal S}_{\mscr{micro}}}
\newcommand{\BHfree}{{\mathcal F}_{\rm BH}}
\newcommand{\BHZ}{{Z_{\rm BH}}}
\newcommand{\Ftop}{{F_{\rm top}}}
\newcommand{\Ztop}{{Z_{\rm top}}}

\usepackage{amsmath}
\usepackage{amssymb}
\usepackage{amscd}
\usepackage{bbm}

\begin{document}

\renewcommand{\thefootnote}{\fnsymbol{footnote}}

\begin{titlepage}
\begin{center}
\hfill LTH 685 \\
\hfill {\tt hep-th/0512048}\\
\vskip 10mm

{\Large
{\bf Strings, higher curvature corrections, and black holes}}\footnote{
Based on a talk given at the 2nd Workshop on Mathematical
and Physical Aspects of Quantum Gravity, Blaubeuren, 28 July -- 
1 August, 2005.} 
\vskip 10mm

\textbf{T. Mohaupt} 

\vskip 4mm

Theoretical Physics Division\\
Department of Mathematical Sciences\\
University of Liverpool\\
Liverpool L69 3BX, UK \\
{\tt Thomas.Mohaupt@liv.ac.uk}
\end{center}

\vskip .2in 

\begin{center} {\bf ABSTRACT} \end{center}
\begin{quotation} \noindent
We review old and recent results on subleading contributions
to black hole entropy in string theory.
\end{quotation}

\vfill

\end{titlepage}

\eject

\renewcommand{\thefootnote}{\arabic{footnote}}

\section{Introduction}
\setcounter{equation}{0}
\setcounter{footnote}{0}

The explanation of black hole entropy in terms of microscopic states
is widely regarded as one of the benchmarks for theories of 
quantum gravity. The analogy between the laws of black hole mechanics
and the laws of thermodynamics, combined with the Hawking effect, suggests
to assign to a black hole of area $A$ the `macroscopic' (or `thermodynamic')
entropy\footnote{We
work in Planckian units, where $c=\hbar=G_N=1$. We also set $k_B =1$.}
\be
\Smacro = \frac{A}{4} \;.
\ee
$\Smacro$ depends on a small number of parameters which can
be measured far away from the black hole and determine its
`macroscopic' state: the mass $M$, the angular momentum $J$ and
its charges $Q$ with respect to long range gauge forces. A
theory of quantum gravity should be able to specify and count
the microstates of the black hole which give rise to the
same macrostate. If there are $N$ states corresponding to
a black hole with parameters $M,J,Q$, then the associated 
`microscopic' or `statistical' entropy is
\be
\Smicro = \log N \;.
\ee
By the analogy to the relation between thermodynamics and statistical
mechanics, it is expected that the macroscopic and microscopic 
entropies agree.\footnote{As we will see later, there are examples
where they agree in leading order of a semiclassical expansion, but
disagree at the subleading level. This is a success rather than a 
problem because the discrepancies can be explained:
the entropies that one compares correspond to different statistical
ensembles.}

The characteristic feature of string theory is the existence
of an infinite tower of excitations with ever-increasing mass. 
Therefore it is natural to take the fundamental strings 
themselves as candidates for the black hole microstates \cite{Sus,HorPol}. 
In the realm of perturbation theory, which describes strings
moving in a flat background space-time at asymptotically small
string coupling one has access to the number of states with a 
given mass. The asymptotic number of states at high mass
is given by the famous formula of Hardy-Ramanujan. Taking the 
open bosonic string for definiteness, the mass formula is
$\alpha' M^2 =  N  -1 $, where $\alpha'$ is the
Regge slope (the only independent dimensionful constant of
string theory)
and $N \in {\mathbbm N}$ is the excitation level. For large
$N$ the number of states grows like $\exp(\sqrt{N})$, so that the
statistical entropy grows like 
\be
\Smicro \approx  \sqrt{N} \;.
\ee
It is clear that when increasing the mass (at finite
coupling), or, alternatively, when increasing the coupling while keeping
the mass fixed, the backreaction of the string onto its ambient
space-time should lead to the formation of a black hole. Roughly,
this happens when the Schwarzschild radius $r_S$ of the string equals
the string length $l_S=\sqrt{\alpha'}$. Using the relation 
$l_P = l_S g_S$
between 
the string length, the Planck length $l_P$ and the dimensionless
string coupling $g_S$, 
together with the fact that a black hole of mass $\alpha' M^2 \approx N$
has Schwarzschild radius $r_S \approx G_N M \approx \sqrt{\alpha'} g_S^2 
\sqrt{N}$ and entropy $\Smacro \approx \ft{A}{G_N} \approx g_S^2 N$
one finds that \cite{Sus,HorPol}\footnote{In this paragraph we have 
reconstructed the
dimensionful quantitites $G_N$ and $\alpha'$ for obvious reasons. All
approximate identities given here hold up to multiplicative constants
of order unity and up to subleading additive corrections.}
\be
r_S \approx l_S \Leftrightarrow g_S^2 \sqrt{N} \approx 1 \;.
\ee
It is precisely in this regime that the 
entropy of string states $\Smicro \approx \sqrt{N}$
equals the entropy $\Smacro \approx g_S^2 N$ of a black hole 
with the same mass, up to factors of
order unity. The resulting scenario of a string -- black hole correspondence,
where strings convert into black holes and vice versa at a threshold
in mass/coupling space is quite appealing. In particular, it applies to 
Schwarzschild-type black holes and makes a proposal for the 
final state of black hole evaporation, namely the conversion into
a highly excited string state of the same mass and entropy. 
However, this picture is very qualitative and one would 
like to have examples where one can make a quantitative 
comparision or even a precision test of the relation between 
macroscopic and microscopic entropy. 

Such examples are available, if one restricts oneself to 
supersymmetric states, also called BPS states. We will consider
four-dimensional black holes, where the setup is
as follows: one considers a string compactification
which preserves a four-dimensional supersymmetry algebra with
central charges $Z$. Then there exist supermultiplets on 
which part of the superalgebra is realized trivially. These multiplets
are smaller than generic massive supermultiplets (hence also called
`short multiplets'), they saturate a mass bound of the form
$M\geq |Z|$, and many of their properties are severely restricted
by the supersymmetry algebra. By counting all states of given mass
and charges, one obtains the statistical entropy $\Smicro$. 
This can now be compared to the entropy $\Smacro$ of a black hole
which has the same mass, carries the same charges and 
is invariant under the same supertransformations. As above, 
the underlying idea
is that by increasing the string coupling we can move
from the regime of perturbation theory in flat space to a regime where
the backreaction onto space-time has led to the formation of a black hole.
This regime can be analyzed by using the low-energy effective field
theory of the massless string modes, which encodes all long range
interactions. The corresponding effective action can be 
constructed using string perturbation theory and is valid 
for small (but finite) coupling $g_S \leq 1$ and for space-time
curvature which is small in units of the string length.
One then constructs supersymmetric black hole solutions with the
appropriate mass and charge. A black hole solution is called 
supersymmetric if it has Killing spinors, which are the 
`fermionic analogues' of Killing vectors. More precisely, if we
denote the supersymmetry transformation parameter by $\epsilon(x)$,
the fields collectively by $\Phi(x)$, and the particular field configuration
under consideration by $\Phi_0(x)$, 
then $\epsilon(x)$ is a 
Killing spinor  and $\Phi_0(x)$ is a supersymmetric (or BPS) configuration, 
if the supersymmetry transformation with parameter $\epsilon(x)$ 
vanishes in the background $\Phi_0(x)$:\footnote{This is completely
analogous to the concept of a Killing vector $\xi(x)$, which
generates an isometry of a given metric $g(x)$, i.e., 
$(L_{\xi(x)} g)(x) =0$.} 
\be
\left.
(\delta_{\epsilon(x)}  \Phi) \right|_{\Phi_0(x)} =0  \;.
\ee
We consider black holes which are asymptotically flat. Therefore it
makes sense to say that a black hole is invariant under `the same'
supertransformations as the corresponding string states.
In practise, the effective action is only known up to a certain
order in $g_S$ and $\alpha'$. Thus we need to require that
the string coupling and the curvature at the event horizon 
are small. This can be achieved by taking the charges and, hence,
the mass, to be large.

Having constructed the black hole solution, we can extract the
area of the event horizon and the entropy $\Smacro$ and then compare
to the result of state counting, which yields $\Smicro$. 
Both quantities are measured in different regimes, and therefore
it is  not clear a priori that the number of states is preserved
when interpolating between them. For BPS states,
there exist only two mechanism which can eliminate or create
them when changing parameters:
(i) at lines of marginal stability a BPS multiplet 
can decay into two or more other BPS multiplets, (ii) 
BPS multiplets can combine into non-BPS multiplets. 
It is not yet clear whether these processes really play 
a role in the context of black hole state counting, but 
in principle one needs to deal with them.
One proposal is that 
the quantity which should be compared to the entropy is not the
state degeneracy itself, but a suitable weighted sum,
a supersymmetric index \cite{OSV,Moore1,Moore2}. 
We will ignore these subtleties here and 
take an `experimental' attitude, by just computing $\Smacro$ and
$\Smicro$ in the appropriate regimes and comparing the results.
In fact, we will see that the agreement is spectacular, and 
extends beyond the leading order.\footnote{Of course, at some
level the question whether the black hole entropy corresponds
to the true state degeneracy or to an index becomes relevant.
However, none of the examples analyzed in \cite{Moore1,Moore2}
appears to be conclusive.} 
In particular we will see
that higher derivative terms in the effective action become
important and that one can discriminate between the naive
area law and Wald's generalized definition of black hole entropy 
for generally covariant theories of gravity with higher derivative
terms. Thus, at least for supersymmetric black holes, one can
make precision tests which confirm that the number of microstates
agrees with the black hole entropy.

Besides fundamental strings, string theory contains other
extended objects, which are also important in accounting
for black hole microstates. One particular subclass are the
D-branes. In string perturbation theory they appear as 
submanifolds of space-time, on which open strings can end. 
In the effective field theory they correspond to black p-brane
solutions, which carry a particular kind of charge,
called Ramond-Ramond charge, 
which is not carried by fundamental strings.
In string compactifications one can put the
spatial directions of p-branes along the compact space 
and thereby obtain black holes in the lower-dimensional space-time.
D-branes gave rise
to the first successful quantitative matching between state counting and
black hole entropy \cite{StrVaf}. Here `quantitative' means that the leading
contribution to $\Smicro$ is precisely $\ft{A}{4}$, i.e, the
prefactor comes out exactly. 

When D-branes and other extended objects enter the game, the
state counting becomes more complicated, but the basic ideas
remain as explained above. Also note that 
instead of a flat background space-time
one can consider other consistent string backgrounds.
In particular one can count the microstates of four-dimensional
black holes which arise in string compactifications on tori,
orbifolds and Calabi-Yau manifolds.

\section{The black hole attractor mechanism}
\setcounter{equation}{0}

In this section we discuss BPS black hole solutions of
four-dimensional $N=2$ supergravity. This is the most general
setup which allows supersymmetric black hole solutions and 
arises in various string compactifications, including 
compactifications of the heterotic string on $K3 \times T^2$ 
and of the type-II superstring on Calabi-Yau threefolds. 
The $N=2$ supergravity multiplet is a supersymmetric version 
of Einstein-Maxwell theory: it contains the graviton, a 
gauge field called the graviphoton, and a doublet of Majorana
gravitini. The extreme Reissner-Nordstrom black hole 
is a solution of this theory and provides the simplest 
example of a supersymmetric black hole \cite{Gib,GibHul}.\footnote{
See for example \cite{Torino} for a pedagogical 
treatment.}

In string compactifications the gravity multiplet is always accompanied
by matter multiplets. The only type of matter which is relevant
for our discussion is the vector multiplet, which contains
a gauge field, a doublet of Majorana spinors, and a complex
scalar. We will consider an arbitrary number $n$ of vector multiplets.
The resulting Lagrangian is quite complicated, but all the
couplings are encoded in a single holomorphic function, called
the prepotential \cite{dWLPSvP,dWvP}. 
In order to understand the structure of the
entropy formula for black holes, we need to review some more details.

First, let us note that the fields which are excited in black hole
solutions are only the bosonic ones. Besides the metric there are
$n$ scalar fields $z^A$, $A=1,\ldots, n$ and $n+1$ gauge fields
$F^I_{\mu \nu}$. The field equations are invariant under 
$Sp( 2n + 2, {\mathbbm R})$ transformations, which generalize
the electric-magnetic duality rotations of the Maxwell theory. 
These act linearly on the gauge fields and rotate the $F^I_{\mu \nu}$
among themselves and into their duals. Electric charges
$q_I$ and magnetic
charges $p^I$ are obtained from flux integrals of the dual field
strength and of the field strength, respectively. They 
form a symplectic vector $(p^I, q_J)$. While the metric is
inert, the action on the scalars is more complicated. However, 
it is possible to find a parametrization of the scalar sector
that exhibits a simple and covariant behaviour under symplectic
transformations. The scalar part
of the Lagrangian is a non-linear sigma-model, and the scalar fields
can be viewed as coordinates on a complex $n$-dimensional manifold 
$M$. 
The geometry of $M$ is restriced by supersymmetry, and the 
resulting geometry is known as `special geometry' \cite{dWvP}. 
In the context of the superconformal calculus, the coupling
of $n$ vector multiplets to Poincar\'e supergravity is constructed
by starting with $n+1$ superconformal vector multipletes, and
imposing suitable gauge conditions which fix the additional 
symmetries. As already mentioned, 
the vector multiplet Lagrangian is encoded in 
a single function, the prepotential,
which depends holomorphically on the 
lowest components of the superconformal vector multiplets. 
A consistent coupling to supergravity further requires the
prepotential to be homogenous of degree 2:
\be
F(\lambda Y^I) = \lambda^2 F(Y^I) \;.
\ee
Here, the complex fields $Y^I$, $I=0,1,\ldots, n$ are 
the lowest components of the superconformal vector multiplets.\footnote{
In the context of black hole solutions, it is convenient to work with
rescaled variables. The fields $Y^I$ used in this paper are
related to the lowest components $X^I$ of the superconformal
vector multiplets by $Y^I = \overline{Z} X^I$, where
$Z = p^I F_I - q_I X^I$ is the central charge. 
Using that $F_I$ is homogenous of degree one has
$F_I(Y) = \overline{Z} F_I(X)$. See \cite{CdWM,CdWKM} for
more details.}
The physical scalar fields $z^A$ are given by the
independent ratios, $z^A = \ft{Y^A}{Y^0}$. Geometrically,
the $Y^I$ are coordinates of a complex cone $C(M)$ over the
scalar manifold $M$.  The existence of a prepotential is 
equivalent to the existence of a holomorphic Lagrangian
immersion of $C(M)$ into the complex
symplectic vector space $\mathbbm{C}^{2n+2}$ \cite{ACD}. 
If $(Y^I, F_J)$ are
coordinates on $\mathbbm{C}^{2n+2}$, then, along the 
immersed $C(M)$, the second half of the coordinates
can be expressed in terms of the first half as 
$F_I = \ft{\der F}{\der Y^I}$, where $F$ is the generating
function of the Lagrangian immersion, i.e., the prepotential.\footnote{
It is assumed here that the immersion is generic,
so that that the $Y^I$ are coordinates on $C(M)$. This 
can always be arranged by applying a  symplectic rotation.}
Under symplectic transformations $(Y^I, F_J)$ transforms as a vector.
Therefore it is convenient to use it instead of $z^A$ to 
parametrize the scalar fields.

Let us now turn to static, spherically symmetric, 
supersymmetric black hole solutions. In $N=2$ supergravity,
which has eight independent real supercharges, one can 
have 8 or 4 or 0 Killing spinors. Solutions with 8 Killing spinors
preserve as many supersymmetries as flat space-time and are
regarded as supersymmetric vacua. Besides $\mathbbm{R}^4$ the
only supersymmetric vacua are $AdS^2 \times S^2$ and 
planar waves \cite{Kow-Glik}.\footnote{Presumably this is 
still true in the presesence of neutral matter and including
higher curvature corrections. In \cite{CdWKM} the most general
{\em stationary} vacuum solution for this case was shown to be 
$AdS^2 \times S^2$.}
Supersymmetric black holes are solutions 
with 4 Killing spinors. Since they preserve half as many
supersymmetries as the vacuum, they are called $\ft12$ BPS
solutions. Since these solutions are asymptotically flat,
the number of Killing spinors doubles if one goes to infinity.

One difference between supersymmetric black holes in theories
with vector multiplets 
and the extreme Reissner-Nordstrom black hole of pure $N=2$
supergravity
is that there are several gauge fields, and therefore
several species of electric and magnetic charges. The other 
difference is that
we now have scalar fields which can have a non-trivial
dependence on the radial coordinate. A black hole
solution is parametrized by the magnetic and electric charges 
$(p^I, q_J)$, which are discrete quantities (by Dirac 
quantization) and by the asymptotic values of the scalar 
fields in the asymptotically flat region, $z^A(\infty)$, which can be 
changed continously. In particular, the mass of a black hole can 
be changed continously by tuning the values of the scalar fields at
infinity. The
area of the horizon and hence $\Smacro$ depends on the charges and 
on the values of the scalar fields at the horizon. If the latter could
be changed continuously, this would be at odds with the intended 
interpretation in terms of state counting. 

What comes to the rescue is the so-called black hole
attractor mechanism \cite{FKS}: if one imposes that the solution is
supersymmetric and regular at the horizon, then the values
of the scalars at the horizon, and also the metric, are determined
in terms of the charges. Thus the scalars flow from arbitrary initial values
at infinity to fixed point values at the horizon. The reason behind this
behaviour is that if the horizon is to be finite then the number 
of Killing spinors must double on the horizon. This fixes the 
geometry of the horizon to be of the form $AdS^2 \times S^2$, with
fixed point values for the scalars. In the notation introduced 
above, the values of the scalar fields can be found from the
following black hole attractor equations \cite{FKS}:
\footnote{We use
the notation of \cite{CdWM}, where the scalar fields
$X^I$ used in \cite{FKS} have been rescaled in the way 
explained above.}
\bea
(Y^I - \overline{Y}^I)_{\rm Horizon} &=& i p^I \;,\nonumber \\
(F_I - \overline{F}_I)_{\rm Horizon} &=& i q_I  \;.
\eea
For a generic prepotential $F$ it is not possible to solve 
this set of equations for the scalar fields in closed form. However, 
explicit solutions have been obtained for many physically 
relevant examples, where either the prepotential is sufficiently
simple, or for non-generic
configurations of the charges (i.e., when switching off some of 
the charges) \cite{BCdWKLM,Shm}.\footnote{Also note that if one can solve 
the attractor equations, then one can also find the solution away from 
the horizon \cite{BLS}.} 

The entropy of the corresponding solution is
\be
\Smacro = \frac{A}{4} = \pi |Z|^2_{\rm Horizon} = \pi
(p^I F_I - q_I Y^I)_{\rm Horizon} \;.
\ee
Here $Z$ is a particular, symplectically invariant contraction of
the fields with the charges, 
which gives the central charge carried by the solution when 
evaluated at infinity. At the horizon, this quantity sets the scale
of the $AdS^2 \times S^2$ space and therefore gives the area $A$.

Let us take a specific example. We consider the prepotential 
of a type-II Calabi-Yau compactification at leading order in 
both the string coupling $g_S$ and the string scale $\alpha'$.
If we set half of the charges to zero, $q_A=p^0=0$, then
the attractor equations can be solved explicitly. To ensure
weak coupling and small curvature at the horizon, the non-vanishing
charges must satisfy $|q_0| \gg p^A \gg 1$.\footnote{In our conventions
$q_0<0$ under the conditions stated in the text.}
The resulting entropy
is \cite{BCdWKLM}:
\be
\Smacro = 2 \pi \sqrt{ \ft16 |q_0| C_{ABC} p^A p^B p^C } \;.
\label{SmacroCY}
\ee
Here $C_{ABC}$ are geometrical parameters (triple intersection 
numbers) which depend on the specific Calabi-Yau threefold used
for compactification.

For this example the state counting has been performed using 
the corresponding brane configuration. The result is \cite{MSW,Vaf}:
\be
\Smicro = 2 \pi \sqrt{ \ft16 |q_0| (C_{ABC} p^A p^B p^C + c_{2A}p^A)} \;,
\label{MSWV}
\ee
where $c_{2A}$ is another set of geometrical parameters of the
underlying Calabi-Yau threefolds (the components of the second
Chern class with respect to a homology basis). Since this
formula contains a subleading term, which is not covered by
the macroscopic entropy (\ref{SmacroCY}), 
this raises the question how one can
improve the treatment of the black hole solutions. 
Since we interpret supergravity actions as effective
actions coming from string theory, the logical 
next step is to investigate the effects of higher 
derivative terms in the effective action, which are induced
by quantum and stringy corrections.

\section{Beyond the area law}
\setcounter{equation}{0}

There is a particular class of higher derivative terms for which
the $N=2$ supergravity action can be constructed explicitly
\cite{deW1} (see also \cite{TMhabil} for a review). 
These terms are encoded in the so-called Weyl multiplet and can
be taken into account by giving the prepotential a dependence
on an additional complex variable $\Upsilon$, which is proportional to 
the lowest
component of the Weyl multiplet. The equations of motion relate
$\Upsilon$ to the (antiselfdual part of the) graviphoton field
strength. The generalized prepotential is holomorphic and
homogenous of degree 2:
\be
F(\lambda Y^I, \lambda^2 \Upsilon) = \lambda^2 F(Y^I, \Upsilon) \;.
\label{HomGen}
\ee
Expanding in $\Upsilon$ as
\be
F(Y^I, \Upsilon) = \sum_{g=0}^\infty F^{(g)}(Y^I) \Upsilon^g \;,
\ee
one gets an infinite sequence of coupling functions $F^{(g)}(Y^I)$.
While $F^{(0)}(Y^I)$ is the prepotential, the $F^{(g)}(Y^I)$, $g\geq 1$,
are coefficients of higher derivative terms. Among these are 
terms of the form
\be
F^{(g)}(Y^I) (C^-_{\mu \nu \rho \sigma})^2 (F^-_{\tau \lambda})^{2g-2}
+ \mbox{ c.c.} \;, 
\ee
where $C^-_{\mu \nu \rho \sigma}$ and $F^-_{\tau \lambda}$ are
the antiselfdual projections of the Weyl tensor and of the 
graviphoton field strength, respectively. In the context
of type-II Calabi-Yau compactifications the functions $F^{(g)}(Y^I)$
can be computed using a topologically twisted version of the theory 
\cite{BCOV,AGNT}.

Starting from the generalized prepotential one can work out the
Lagrangian and construct static, spherically symmetric BPS
black hole solutions \cite{CdWM,CdWKM}.\footnote{In fact, 
one can construct stationary
BPS solutions which generalize the IWP solutions of pure 
supergravity \cite{CdWKM}.}
It can be shown that the near horizon solution is still determined
by the black hole attractor equations,\footnote{More precisely, the
attractor equations are necessary and sufficient for having a 
fully supersymmetric solution with 8 Killing spinors at the horizon.
The geometry is still $AdS^2 \times S^2$, but with a modified scale.}
 which now involve the generalized
prepotential \cite{CdWM}:
\bea
(Y^I - \overline{Y}^I)_{\rm horizon} &=& i p^I  \;,\nonumber \\
(F_I(Y,\Upsilon) - \overline{F}_I(\overline{Y}, 
\overline{Y}))_{\rm horizon} &=& i q_I \;.
\label{AttrEqsR2}
\eea
The additional variable $\Upsilon$ takes the value $\Upsilon=-64$ 
at the horizon. Since the generalized prepotential enters into the
attractor equations, the area of the horizon is modified by the
higher derivative terms. Moreover, there is a second modification, 
which concerns the very definition of the black hole entropy.

The central  argument for interpreting the area of the horizon as an entropy
comes from the first law of black hole mechanics, which relates
the change of the mass of a stationary black hole 
to changes of the area and of other quantities
(angular momentum, charges):
\be
\delta M = \frac{\kappa_S}{8 \pi} \delta A + \cdots  \;,
\ee
where $\kappa_S$ is the surface gravity of the black hole.\footnote{See
for example \cite{TMhabil} for a review of the relevant properties
of black hole horizons.}
Comparing to the first law of thermodynamics,
\be
\delta U  = T \delta S + \cdots  \;,
\ee
and taking into account that the Hawking temperature of a black hole
is $T = \frac{\kappa_S}{2 \pi}$, one is led to the identification
$\Smacro = \ft{A}{4}$. This is at least the situation in 
Einstein gravity. The first law can be generalized to more general
gravitational Lagrangians, which contain higher derivative terms,
in particular arbitrary powers of the Riemann tensor and of its 
derivatives \cite{Wald1,Wald2}.
The basic assumptions entering the derivation
are that the Lagrangian is generally covariant, and that it admits 
stationary black hole solutions whose horizons are Killing horizons.
Then there still is a first law of the form
\be 
\delta M =  \frac{\kappa_S}{2 \pi} \delta \Smacro + \ldots \;,
\ee
but $\Smacro \not= \ft{A}4$ in general. Rather, $\Smacro$ is given
by the surface charge associated with the horizontal Killing vector
field $\xi$:
\be
\Smacro = 2 \pi \oint_{\rm horizon} {\bf Q}[\xi] \;,
\ee
which can be expressed in terms of variational derivatives of the
Lagrangian with respect to the Riemann tensor \cite{Wald2}:
\be
\Smacro = - 2 \pi \oint_{\rm horizon} \frac{\delta L}{\delta R_{\mu \nu \rho
\sigma}} \varepsilon_{\mu \nu} \varepsilon_{\rho \sigma} \;
\sqrt{h} \; d^2 \theta \;.
\label{WaldGen}
\ee
Here $\varepsilon_{\mu \nu}$ is the normal bivector of the
horizon, normalized as $\varepsilon_{\mu \nu} 
\varepsilon^{\mu \nu} =-2$, and $h$ is the pullback of the metric 
onto the horizon.\footnote{In carrying out the variational derivatives
one treats the Riemann tensor formally as if it was independent of 
the metric. At first glance this rule looks ambigous, because one can
perform partial integrations. But the underlying formalism guarantees
that the integrated quantity $\Smacro$ is well defined \cite{Wald1,Wald2}
(see also \cite{CdWM99} for an alternative proof).}
From (\ref{WaldGen}) it is clear that corrections to the area law
will be additive:
\be
\Smacro = \frac{A}{4} + \cdots \;.
\ee
Here the leading term comes from the variation of the Einstein-Hilbert
action. 

The general formula (\ref{WaldGen}) can be evaluated for
the special case of $N=2$ supergravity with vector multiplets and
higher derivative terms encoded in the generalized prepotential.
The result is \cite{CdWM}:
\be
\Smacro(q,p) = \pi \left( (p^I F_I - q_I Y^I) + 4 \;
\mbox{Im}( \Upsilon F_{\Upsilon}) \right)_{\rm horizon} \;,
\label{WaldSpec}
\ee
where $F_{\Upsilon} = \ft{\der F}{\der \Upsilon}$. Thus $F_\Upsilon$,
which depends on the higher derivative couplings $F^{(g)}$, $g\geq 1$, 
encodes the corrections to the area law.

If the prepotential is sufficiently simple, one can find
explicit solutions of the attractor equations 
\cite{CdWM,CdWM99,CdWM9906}.\footnote{One can also construct the
solution away from the horizon, at least iteratively 
\cite{CdWKM,CdWKM0012}.}
In particular, we can now compare $\Smacro$ to the $\Smicro$ computed from
state counting (\ref{MSWV}) \cite{CdWM}:
\bea
&& \Smacro(q,p) \nonumber \\
& & = \frac{A}{4} + \mbox{Correction term} \nonumber \\
 & & = 
2 \pi \frac{
\ft16 |q_0|( C_{ABC} p^A p^B p^C + \ft12 c_{2A}p^A )
}
{
\sqrt{
\ft16 |q_0| (C_{ABC} p^A p^B p^C + c_{2A} p^A)
}
} 
+ 2 \pi 
\frac{ 
\ft1{12} |q_0| c_{2A} p^A 
}{
\ft16 |q_0| (C_{ABC} p^A p^B p^C + c_{2A} p^A)
}
    \nonumber \\
&&= 2 \pi \sqrt{\ft16 |q_0| (C_{ABC} p^A p^B p^C + c_{2A} p^A)} 
\nonumber \\
&&=\Smicro \;.
\label{MacroMicroCY3}
\eea
In the second line we can see explicitly how the higher derivative
terms modify the area. But, when sticking to the naive area law,
one finds that $\ft{A}{4}$ differs from $\Smicro$ already
in the first subleading term in an expansion in large charges.
In contrast, when taking into account the modification of the
area law, $\Smacro$ and $\Smicro$ agree completely. In other
words `string theory state counting knows about the modification
of the area law.'  This provides strong evidence that string theory
captures the microscopic degrees of freedom of black holes, at least
of supersymmetric ones.

At this point one might wonder about the role of other types
of higher derivatives terms. So far, we have only included a
very particular class, namely those which can be described 
using the Weyl multiplet. The full string effective action
also contains other higher derivative terms,
including terms which are higher powers in the curvature.
Naively, one would expect that these also contribute
to the black hole entropy. However, as we will see 
in the next sections, one can obtain an even more impressive
agreement between microscopic and macroscopic entropy by
just using the terms encoded in the Weyl multiplet. One reason
might be the close relationship between the terms described by
the Weyl multiplet and the topological string, which we are
going to review in the next section. There are two other
observation which indicate that the Weyl multiplet encodes
all contributions relevant for the entropy.\footnote{For 
toroidal compactifications of type-II string theory there are
no $R^2$-corrections, but the entropy of string states is
non-vanishing. This case seems to require the 
presence of higher derivative terms which are not 
captured by the Weyl multiplet. See \cite{Moore1} for
further discussion.}
The first observation is that when one just adds 
a Gauss-Bonnet term to the Einstein Hilbert action, one
obtains the same entropy formula (\ref{MSWV}) as when using
the full Weyl multiplet \cite{BCM,Sen:0508}. The second is
that (\ref{MSWV}) can also be derived using gravitational
anomalies \cite{HarMinMoo,KraLar}. Both suggest that the
black hole entropy is a robust object, in the sense that
it does not seem to depend sensitively on details of the
Lagrangian. 

One might also wonder, to which extent the matching of
microscopic and macroscopic entropy 
depends on
supersymmetry. Here it is encouraging that the derivations
of (\ref{MSWV}) in \cite{Sen:0506,Sen:0508} and 
\cite{KraLar} do not invoke  supersymmetry directly. 
Rather,  \cite{Sen:0506,Sen:0508} analyses 
black holes with near horizon geometry $AdS^2 \times S^2$
in the context of general higher derivative covariant actions,
without assuming any other specifics of the
interactions. This leads to a formalism based on an entropy
function, which is very similar to the one found for
supersymmetric black holes some time ago \cite{BCdWKLM}, and which
we will review in a later section. The work of \cite{KraLar}
relates Wald's entropy formula to the AdS/CFT correspondence.

Finally, it is worth remarking that according to 
\cite{Sen:0506,KraLar} similar results should hold in 
space-time dimensions other than four. A particularly 
interesting dimension seems to be five, because
there is a very close relationship between 
four-dimensional supersymmetric black holes and
five-dimensional supersymmetric rotating black holes
and black rings \cite{GHLS,BCM}, which holds in 
the presence of higher curvature terms.

Coming back to (\ref{MacroMicroCY3}), we remark that it is 
intriguing that two 
complicated terms, the area and the correction term, combine
into a much simpler expression. This suggests that, 
although (\ref{WaldSpec}) is a sum of two terms, it should be 
possible to express the entropy in terms of one single function.
Though it is not quite obvious how to do this, it is
in fact true.

\section{From black holes to topological strings}
\setcounter{equation}{0}

The black hole entropy (\ref{WaldSpec}) can be written as the
Legendre transform of another function ${\mathcal F}_{\rm BH}$, which
is interpreted as the black hole free energy. This is seen as follows
\cite{OSV}.
The `magnetic' attractor equation $Y^I - \overline{Y}^I =ip^I $ can 
be `solved' by setting:\footnote{We use the notation of \cite{CdWM}
which is slightly different from the one of \cite{OSV}.}
\be
Y^I = \frac{\phi^I}{2 \pi} + i \frac{p^I}{2} \;,
\ee
where $\phi^I \propto {\rm Re}Y^I $ is determined by
the remaining `electric' attractor equation. From the
gauge field equations of motion in a stationary space-time
one sees that $\phi^I$ is proportional to the electrostatic
potential (see for example \cite{CdWKM}). Now define the free energy
\be
{\mathcal F}_{\rm BH}(\phi, p) := 4 \pi 
\mbox{Im} F(Y, \Upsilon)_{\rm horizon} \;.
\ee
Observe that the electric attractor equations $F_I - \overline{F}_I
= ip^I$ are equivalent to
\be
\frac{\der \BHfree}{\der \phi^I} = q_I \;.
\ee
Next note that the homogenity property (\ref{HomGen}) of the 
generalized prepotential implies the Euler-type relation
\be
2 F = Y^I F_I + 2 \Upsilon F_\Upsilon \;.
\ee
Using this one easily verifies that
\be
\BHfree(\phi, p) - \phi^I \frac{\der \BHfree}{\der \phi^I} =
\Smacro(q,p) \;.
\label{FreeEnSmacro}
\ee
Thus the black hole entropy is obtained from the black hole
free energy by a (partial) Legendre transform which replaces the electric
charges $q_I$ by the electrostatic potentials $\phi^I$. 

This observation opens up various routes of investigation. Let us
first explore the consequences for the relation between 
$\Smacro$ and $\Smicro$. The black hole partition function 
associated with $\BHfree$ is 
\be 
\BHZ(\phi,p) = e^{\BHfree (\phi,p)}  \;.
\ee
Since it  depends on $\phi^I$ rather than on $q_I$, it is clear
that this is not a microcanonical partition function. Rather it
refers to a mixed ensemble, where the magnetic charges have been
fixed while the electric charges fluctuate. The electrostatic 
potential $\phi^I$ is the corresponding thermodynamic potential.
However, the actual state degeneracy $d(q,p)$
should be computed in the
microcanonical ensemble, where both electric and magnetic charges 
are fixed. Using a standard thermodynamical relation, we see
that $\BHZ$ and $d(q,p)$ are formally related by a (discrete)
Laplace transform:
\be
Z_{\rm BH}(\phi,p) = \sum_q d(q,p) e^{\phi^I q_I} \;.
\label{DiscLapl}
\ee
We can solve this formally for the state degeneracy by
an (inverse discrete) Laplace transform,
\be
d(q,p) = \int d \phi \; e^{ \BHfree(\phi,p) - \phi^I q_I }
\label{MicroDeg}
\ee
and express the microscopic entropy
\be
\Smicro(q,p) = \log d(q,p)
\ee
in terms of the black hole free energy. Comparing 
(\ref{FreeEnSmacro}) to (\ref{MicroDeg}) it is
clear that
$\Smacro$ and $\Smicro$ will not be equal in general. Both can be
expressed in terms of the free energy, but one is given
through a Laplace transform and the other through a Legendre
transform \cite{OSV}. From statistical mechanics we are used to the fact
that quantities might differ when computed using different ensembles, 
but we expect
them to agree in the thermodynamic limit. In our context the
thermodynamic limit corresponds to the limit of large charges, in which
it makes sense to evaluate the inverse Laplace transform (\ref{MicroDeg})
in a saddle point approximation:
\be
e^{\Smicro(q,p)} = \int d \phi e^{\BHfree(\phi, p) - \phi^I q_I}
\approx e^{\BHfree(\phi,p) - \phi^I \frac{\der \BHfree}{\der \phi^I}}
= e^{\Smacro(q,p)} \;.
\ee
Since the saddle point value of the inverse Laplace transform
is given by the Legendre transform, we see that both entropies
agree in the limit of large charges. Note that already the first
subleading correction, which comes from quadratic fluctuations
around the saddle point, will in general lead to deviations.
We will illustrate the
relation between $\Smacro$ and $\Smicro$ using specific examples later on.

We now turn to another important consequence (\ref{FreeEnSmacro}).
As already mentioned the couplings $F^{(g)}(Y)$ of the effective
$N=2$ supergravity Lagrangian can be computed within the topologically
twisted version of type-II string theory with the relevant Calabi-Yau
threefold as target space. The effect of the topological twist is roughly
to remove all the non-BPS states, thus reducing each charge sector
to its ground state.
The coupling functions can be encoded
in a generating function, called the topological free energy 
$F_{\rm top}(Y^I, \Upsilon)$, which equals the generalized 
prepotential $F(Y^I, \Upsilon)$
of supergravity up to a conventional overall
constant. The associated topological partition function
\be
\Ztop = e^{\Ftop}
\ee
can be viewed as a partition function for the BPS states
of the full string theory. Taking into account the conventional
normalization factor between $\Ftop$ and $F(Y^I,\Upsilon)$ one
observes \cite{OSV}:
\be
\BHZ = e^{\BHfree} = e^{4 \pi {\rm Im} F} = e^{\Ftop + 
\overline{\Ftop}} = | e^\Ftop |^2 = |\Ztop|^2 \;.
\label{ZBH-Ztop}
\ee
Thus there is a direct relation between the black hole entropy
and the topological partition function, which suggests that the
matching between macroscopic and microscopic entropy extends far
beyond the leading contributions. Moreover, the relation
$\BHZ = |\Ztop|^2$ suggests to interprete $\Ztop$ as a quantum mechanical
wave function and $\BHZ$ as the associated probability \cite{OSV}. This can be
made precise as follows: $\Ztop$ is a function on the vector multiplet
scalar manifold, which in type-IIA (type-IIB) Calabi-Yau compactifications
coincides with the moduli space of complexified K\"ahler structures
(complex structures). This manifold is in particular symplectic, and 
can be interpreted as a classical phase space. Applying geometric quantization
one sees that $\Ztop$ is indeed a wave function on the resulting Hilbert
space \cite{WittenBGI}. This reminds one of the minisuperspace
approximations used in canonical quantum gravity. In our case the
truncation of degrees of freedom is due to the topological twist, 
which leaves the moduli of the internal manifold as the remaining
degrees of freedom. In other words the full string theory is 
reduced to quantum mechanics on the moduli space. One is not
restricted to only discussing black holes in this framework, 
but, by a change of perspective and some modifications, one can approach
the dynamics of flux compactitications and quantum cosmology \cite{OVV}.

The link between black holes and flux compactifications is 
provided by the observation that from the higher-dimensional
point of view the near-horizon geometry of a supersymmetric
black hole is $AdS^2 \times S^2 \times X^*$, where $X^*$ denotes the 
Calabi-Yau threefold at the attractor point in moduli space corresponding
to the charges $(q^I, p_I)$. This can be viewed as a flux compactification
to two dimensions.  The flux is given 
by the electric and magnetic fields along $AdS^2 \times S^2$,
which are covariantly constant, and compensate for the fact that
the geometry is not Ricci-flat. From the two-dimensional perspective
the attractor mechanism reflects that the reduction on 
$S^2$ gives rise to a gauged supergravity theory with a nontrivial 
scalar potential which fixes the moduli. When taking the 
spatial direction of $AdS^2$ to be
compact, so that space takes the form $S^1 \times S^2 \times X^*$,
then vacua with different
moduli are separated by barriers of finite energy. As a consequence,
the moduli, which otherwise label superselection sectors, can fluctuate.
In this context 
$\Ztop$ has been interpreted as a Hartle-Hawking type wave function
for flux compactifications \cite{OVV}, while \cite{GukSarVaf}
argued that string compactifications with asymptotically 
free gauge groups are preferred.

It should be stressed that there are many open questions
concerning these proposals, both conceptually and technically.
Some of these will be discussed in the next section from
the point of view of supergravity and black holes. Nevertheless
these ideas are very interesting because they provide a new
way to approach the vacuum selection problem of string theory.
Moreover there seems to be a lot in common with the canonical approach
to quantum gravity and quantum cosmology. This might help to develop
new ideas how to overcome the shortcomings of present day string 
theory concerning time-dependent backgrounds. By phrasing
string theory in the language used in canonical quantum gravity,
one would have a better basis for debating the merits of 
different approaches to quantum gravity.

\section{Variational principles for black holes}
\setcounter{equation}{0}

We will now discuss open problems concerning the formulae 
(\ref{DiscLapl}), (\ref{MicroDeg}) and (\ref{ZBH-Ztop}) which
relate the black hole entropy to the counting of microstates.
The following sections are based on \cite{CdWKM05}.\footnote{Preliminary 
results have already been presented at conferences, including 
the `Workshop on gravitational aspects of string theory'
at the Fields Institute (Toronto, May 2005) and
the `Strings 2005' (Toronto, July 2005). See
{\tt http://online.kitp.ucsb.edu/online/joint98/kaeppeli/} and \\
{\tt http://www.fields.utoronto.ca/audio/05-06/strings/wit/index.html}.} 
See also \cite{B1,B2,J} for further discussion.

Consider for definiteness (\ref{MicroDeg}):
\be
d(q,p) = \int d \phi \; e^{ \BHfree(\phi,p) - \phi^I q_I } \;,
\ee
which relates the black hole free energy to the microscopic state
degeneracy. This is formally an inverse discrete Laplace transformation,
but without specifying the integration contour it is not clear
that the integral converges. We will not address this issue here, but
treat the integral as a formal expression which we evaluate 
asymptotically by saddle point methods. The next issue is the
precise form of the integrand. As we stressed above various quantities
of the effective $N=2$ supergravity, in particular the charges 
$(p^I, q_J)$, are subject to symplectic rotations. The microscopic
state degeneracy $d(q,p)$ is an observable and therefore should
transform covariantly, i.e., it should be a symplectic function.
Moreover, string theory has discrete symmetries, in particular
T-duality and S-duality, which are realized as specific symplectic
transformations in $N \geq 2$ compactifications. Since these are
symmetries, $d(q,p)$ should be invariant under them. The transformation
properties with respect to symplectic transformations are simple and
transparent as long as one works with symplectic vectors, such
as $(Y^I, F_J)$ and its real and imaginary parts. By the Legendre
transform we now take $\phi^I$ and $p^I$ as our independent 
variables, and these do not form a symplectic vector. Thus manifest
symplectic covariance, as it is present in the entropy formula
(\ref{WaldSpec}), has been lost. Moreover, it is clear that
if $d\phi$ is the standard Euclidean measure $\prod_{I} d \phi^I$, 
then the integral cannnot be expected to be symplectically invariant. 
From the point of view
of symplectic covariance one should expect that the integration measure
is symplectically invariant, while the integrand is a symplectic function.
We will now outline a systematic procedure which provides a modified 
version of (\ref{MicroDeg}) which has this property. 

The starting point is the observation that the entropy of supersymmetric
black holes can be obtained from a variational principle 
\cite{BCdWKLM,CdWKM05}. Define
the symplectic function
\be
\Sigma (q,p,Y,\overline{Y}) := - K - W - \overline{W} 
+ 128 i F_\Upsilon - 128 i \overline{F}_{\overline{\Upsilon}} \;,
\ee
where 
\be
K := i(\overline{Y}^I F_I - \overline{F}_I Y^I) \;\;\;
\mbox{and} \;\;\;
W := q_I Y^I - p^I F_I \;.
\ee
One then finds that the conditions for critical points of $\Sigma$,
\be
\frac{\der \Sigma}{\der Y^I} = 0 =
\frac{\der \Sigma}{\der \overline{Y}^I} 
\ee
are precisely the attractor equations (\ref{AttrEqsR2}).
Moreover, at the attractor we find that
\be
\pi \Sigma_{\rm attractor}(q,p) = \Smacro(q,p)  \;.
\ee
We also note that one can split the extremization procedure consistently
into two steps. If one first extremizes $\Sigma$ with respect to the
imaginary part of $Y^I$, one obtains the magnetic attractor equations.
Plugging these back we find 
\be
\pi \Sigma(\phi,q,p)_{\mscr{magnetic attractor}}  =
\BHfree(\phi,p) - \phi^I q_I \;
\ee
and recover the free energy of \cite{OSV} at an intermediate
level.
Subsequent extremization with respect to $\phi^I \propto \mbox{Re} Y^I$
gives the electric attractor equations, and by plugging them
back we find the entropy. Moreover, while the
free energy ${\mathcal F}_{BH}(\phi,p)$ is related to the black hole
entropy ${\mathcal S}_{\rm macro}(p,q)$
by a partial Legendre transform, the charge-independent part
of $\Sigma$, namely $-K + 128 i F_\Upsilon
- 128 i \overline{F}_{\overline{\Upsilon}}$ is its full Legendre
transform.

Since $\pi \Sigma (q,p,Y,\overline{Y})$ is a symplectic function, 
which equals $\Smacro$ at its critical point, it is natural 
to take $\exp( \pi \Sigma)$ to define a modified version of 
(\ref{MicroDeg}). This means that we should not only to integrate over 
$\phi^I \propto \mbox{Re} Y^I$, but also over the other scalar fields
$\mbox{Im}Y^I$.
What about the measure? Since it should be symplectically invariant,
the natural choice is\footnote{This follows from inspection of the
symplectic transformation rules \cite{deW1}. 
Alternatively, one might note that
this measure is proportional to the top exterior power of the
natural symplectic form of $C(M)$, the cone over the 
moduli space.}
\be
d \mu(Y,\overline{Y}) = \prod_{IJ} dY^I d \overline{Y}^J 
\det ( -2i \mbox{Im} (F_{KL}) ) \;,  
\ee
where $F_{KL}$ denotes the second derivatives of the generalized
prepotential with respect to the scalar fields.
Putting everything together the  proposal of \cite{CdWKM05} 
for a modified version of (\ref{MicroDeg}) is:
\be
d(q,p) = {\mathcal N} \int \prod_{IJ} dY^I d \overline{Y}^J 
\det ( -2i \mbox{Im} (F_{KL}) )  \exp \left(
\pi \Sigma \right)  \;,
\label{MicroDegOurs}
\ee
where ${\mathcal N}$ is a normalization factor.
In order to compare to (\ref{MicroDeg}) it is useful to note that
one can perform the saddle point evaluation in two steps. In the first 
step one takes a saddle point with respect to 
the imaginary parts of $Y^I$, which
imposes the magnetic attractor equations. Performing the saddle
point integration one obtains
\be
d(q,p) = {\mathcal N}' \int \prod_I d \phi^I 
\sqrt{ \det ( -2i (\mbox{Im} F_{KL})_{\mscr{magn. attractor}} ) } \exp \left(
\BHfree - \phi^I q_I \right) \;,
\ee
which is similar to the original 
(\ref{MicroDeg}) but contains a non-trivial measure factor 
steming from the requirement of symplectic covariance.
Subsequent saddle point evaluation 
with respect to $\phi^I\propto \mbox{Re} Y^I$ gives
\be
d(q,p) = \exp \left( \pi \Sigma_{\rm attractor} \right)
= \exp \Smacro \;.
\ee

Let us next comment on another issue concerning 
(\ref{MicroDeg}). So far we have been working with a holomorphic 
prepotential $F(Y^I, \Upsilon)$, which upon differentiation yields
effective gauge couplings that are holomorphic functions of the 
moduli. However, it is well known that the physical couplings
extracted from string scattering amplitudes are not holomorphic.
This can be understood purely in terms of field theory
(see for example \cite{KapLou} for a review): if a 
theory contains massless particles, then the (quantum) effective
action (the generating functional of 1PI Greens function) will in
general be non-local. In the case of supersymmetric gauge theories,
this goes hand in hand with non-holomorphic contributions to gauge 
couplings, which, in $N=2$ theories, cannot be expressed in terms
of a holomorphic prepotential \cite{dWKLL,dWCLMR,CdWM9906}. 
Symmetries, such as S- and T-duality
provide an efficient way of controling these non-holomorphic terms.
While the holomorphic couplings derived from the holomorphic 
prepotential are not consistent with S- and T-duality, the additional
non-holomorphic contributions transform in such a way that 
they restore these symmetries. The same remark applies to the black hole
entropy, as has been shown for the particular case of supersymmetric 
black holes in string compactifications with $N=4$ supersymmetry. 
There, S-duality is supposed to be an exact symmetry, and therefore
physical quantities like gauge couplings, gravitational couplings and 
the entropy must be S-duality invariant. But this is only possible
if non-holomorphic contributions are taken into account 
\cite{HarMor,CdWM9906}. In the notation 
used here this amounts to modifying the black hole free energy 
by adding a real-valued function $\Omega$, which is homogenous of
degree 2 and not harmonic \cite{CdWKM04,CdWKM05}:
\be
\BHfree \rightarrow \wh{\mathcal F}_{\mscr{BH}} 
= 4 \pi \mbox{Im} F(Y^I, \Upsilon)
+ 4 \pi \Omega(Y^I, \overline{Y}^I, \Upsilon, \overline{\Upsilon}) \;.
\ee
Non-holomorphic terms can also be studied in the framework of
topological string theory and are then encoded in a holomorphic
anomaly equation \cite{BCOV}. The role of non-holomorphic
contributions has recently received considerable attention 
\cite{Ver,Sen3,Sen4,Moore1,Moore2} (see also \cite{GS}). It appears that
these proposals do not fully agree with the one of
\cite{CdWKM05}, which was explained above. One way to clarify
the role of non-holomorphic corrections is the study of subleading
terms in explicit examples, which will be discussed in the next
sections.

In the last section we briefly explained how black hole 
solutions are related to flux compactifications. It is interesting
to note that variational principles are another feature that
they share.  Over the last
years it has been realized that the geometries featuring in 
flux compactifications are calibrated geometries, i.e., one
can compute volumes of submanifolds by integrating suitable
calibrating forms over them, without knowing the metric 
explicitly (see for example \cite{GS,DGNV}).
Such geometries can be characterized in terms of variational 
principles, such as Hitchin's \cite{Hitchin}. In physical 
terms the idea is to write down an abelian gauge theory
for higher rank gauge fields (aka differential forms) such
that the equations of motion are the equations characterising
the geometry. The topological partition function $Z_{\rm top}$
should then be interpreted as a wave function of the quantized
version of this theory. Conversely the variational principle
provides the semiclassical approximation of the quantum 
mechanics on the moduli space.

\section{Fundamental strings and `small' black holes}
\setcounter{equation}{0}

So far our discussion was quite abstract and in parts formal. 
Therefore we now want to test these ideas in concrete models. 
As was first realized in \cite{Dab}, the $\ft12$ BPS states of
the toroidally compactified heterotic string provide an ideal
test ground for the idea that there is an exact relation between
black hole microstates and the string partition function. Since
this compactification has $N=4$ rather than $N=2$ supersymmetry,
one gets an enhanced control over both $\Smacro$ and $\Smicro$.
For generic moduli, 
the massless spectrum of the heterotic string compactified on 
$T^6$ consists of the $N=4$ supergravity multiplet together with
22 $N=4$ vector multiplets. Since the gravity multiplet contains
4 graviphotons, the gauge group is $G=U(1)^{28}$. There are 28 electric 
charges $q$ and 28 
magnetic charges $p$ which take values in the Narain lattice
$\Gamma_{6,22}$. This lattice is even and selfdual with respect
to the bilinear form of signature $(6,22)$, and hence it is 
unique up to isometries. The T-duality group $O(6,22,\mathbbm{Z})$
group consists of those isometries which are lattice automorphisms.
The S-duality group $SL(2,\mathbbm{Z})$ acts as ${\bf 2} \otimes 
\mathbbm{1}$ on the $(28 + 28)$-component vector $(q,p) \in 
\Gamma_{6,22} \oplus \Gamma_{6,22} \simeq \mathbbm{Z}^2 \otimes
\Gamma_{6,22}$, where
${\bf 2}$ denotes the fundamental representation of $SL(2,\mathbbm{Z})$
and $\mathbbm{1}$ the identity map on $\Gamma_{6,22}$.

It turns out that the $N=2$ formalism described earlier can be used
to construct supersymmetric black hole solutions of the $N=4$
theory \cite{CdWM9906}. If one uses the up-to-two-derivatives part of the 
effective Lagrangian, the entropy is given by \cite{Cvetic,CdWM9906}
\be
\Smacro = \frac{A}{4} = \pi \sqrt{ q^2 p^2 - (q \cdot p)^2 } \;.
\label{N=4tree}
\ee
Here $a \cdot b$ denotes the scalar product with signature
$(6,22)$, so that the above formula is manifestly invariant under
T-duality. It can be shown that 
$(q^2, p^2, q \cdot p)$ transforms in the ${\bf 3}$-representation
under S-duality, and that the quadratic form 
$q^2 p^2 - (q \cdot p)^2$ is invariant.\footnote{Note that 
$SL(2,\mathbbm{R}) \simeq SO(1,2)$ and that the quadratic form
$q^2 p^2 - (q \cdot p)^2$ has signature $(1,2)$.}
Therefore $\Smacro$ is manifestly S-duality invariant as well. 
The supersymmetric black hole solutions form two classes,
corresponding to the two possible types of BPS multiplets
\cite{SchwSen,DLR} (see also \cite{CCLMR} for a review).
The $\ft12$ BPS solutions with 8 (out of a maximum of 16) Killing spinors 
are characterized by
\be
q^2 p^2 - (q \cdot p)^2 =0
\label{short}
\ee 
and therefore have a degenerate horizon, at least in the lowest
order approximation. Particular solutions of (\ref{short})
are $p=0$ (`electric black holes') and $q=0$ (`magnetic black holes').

The $N=4$ theory also has $\ft14$ BPS solutions with only 4 
Killing spinors. They satisfy 
\be
q^2 p^2 - (q \cdot p)^2 \not=0
\label{intermediate}
\ee 
and therefore have a non-vanishing horizon. They are `genuinely 
dyonic' in the sense that it is not possible to set all electric
(or all magnetic) charges to zero by an S-duality transformation.
Thus they are referred to as dyonic black holes. We will
discuss them in the next section.

Let us return to the electric black holes. We saw that the horizon 
area is zero, $A=0$, and so is the black hole entropy $\Smacro=0$.
Geometrically, the solution has a null singularity, i.e., the
curvature singularity concides with the horizon. One might wonder
whether stringy or quantum corrections resolve this singularity.
Moreover, a vanishing black hole entropy means that there is 
only one microstate, and one should check whether this is true. 

The candidate microstates for the supersymmetric electric black hole
are fundamental strings sitting in $\ft12$ BPS multiplets
\cite{DH}. 
These are precisely the states where the left-moving,
supersymmetric sector is put into its ground state while 
exciting oscillations in the right-moving, bosonic sector. 
Such states take the following form:
\be
\prod_l \alpha^{i_l}_{-m_l}
| (P_L, P_R) \rangle \otimes | {\bf 8} \oplus {\bf 8} 
\rangle
\ee
Here the $\alpha$'s are creation operators for the 
right-moving oscillation 
mode of level $m_l = 1,2,\ldots$, which can be aligned along
the two transverse space directions, $i_l =1,2$, along
the six directions of the torus, $i_l=3, \ldots 8$, or
along the maximal torus of the rank 16 gauge group of the ten-dimensional 
theory, $i_l = 9, \ldots, 24$. $P_L$ and $P_R$ are the left- and
right-moving momenta, or, in other words, $q=(P_L,P_R) \in 
\Gamma_{6,22}$ are the electric charges. Finally $| {\bf 8} \oplus {\bf 8} 
\rangle$ is the left-moving ground state, which is a four-dimensional 
$N=4$  vector supermultiplet with eight bosonic and eight fermionic
degrees of freedom. Since the space-time supercharges 
are constructed out of the left-moving oscillators it is clear
that this state transforms in the same way as the left-moving ground
state, and therefore is an $\ft12$ BPS state.

Physical states satisfy the mass formula 
\be
\alpha' M^2 = N - 1 + \ft12 P_R^2 + \tilde{N} + \ft12 P_L^2 \;,
\ee
where $N, \tilde{N}$ is the total right- and left-moving excitation 
level, respectively. Moreover physical states satisfy the
level matching condition
\be
N - 1 + \ft12 P_R^2 =  \tilde{N} + \ft12 P_L^2 \;.
\ee
For $\ft12$ BPS states we have $\tilde{N}=0$ and level matching fixes
the level $N$ and the mass $M$ in terms of the charges:\footnote{With our 
conventions
$q^2$ is negative for physical BPS states with large excitation number $N$.}
\be
N = \ft12 (P_L^2 - P_R^2) + 1 = - \ft12 q^2 + 1 =
\ft12 |q^2| + 1 \;.
\ee
The problem of counting the number of $\ft12$ BPS states 
amounts to counting partitions of an integer $N$ (modulo
the 24-fold extra degeneracy introduced by the additional labels
$i_l$). This is a classical problem which has been studied
by Hardy and Ramanujan \cite{HR}. The number $d(q)$ of states at of given
charge admits the integral representation
\be
d(q) = \oint_C d \tau \frac{\exp ( i \pi \tau q^2 )}{\eta^{24}(\tau)} \;.
\label{ElectricStates}
\ee
Here $\tau$ take values in the upper half plane, $\eta(\tau)$
is the Dedekind $\eta$-function and $C$ is a suitable integration contour.
For large charges ($|q^2| \gg 1$), 
the asymptotic number of states
is governed by the Hardy-Ramanujan formula
\be
d(q) = \exp \left(
4 \pi \sqrt{ \ft12 |q^2|} - \ft{27}{4} \log |q^2| + \cdots \right) \;.
\ee
The statistical entropy of string states therefore is
\be
\Smicro(q) = 
4 \pi \sqrt{\ft12 |q^2| } - \ft{27}{4} \log |q^2| + \cdots \;.
\ee

Comparing to the black hole entropy of electric supersymmetric
black holes, we realize that there is a discrepancy, because
$\Smacro(q) =0$. One can now reanalyze the black hole solutions
and take into account higher derivative terms. As a first step
one takes those terms which occur at tree level
in the heterotic string theory. These are the same terms that
one gets by dimensional reduction of higher derivative terms
(`$R^4$-terms') of the ten-dimensional effective theory. Already
this leading order higher derivative term 
is sufficient to resolve the null singularity and
to give the electric black hole a finite horizon with area
\be
\frac{A}{4} = 2 \pi \sqrt{ \ft12  |q^2| } = \ft12 \Smicro + \cdots \;,
\ee
as was shown in \cite{DKM} using the results of \cite{CdWM,CdWM9906}.
The resulting area is large in Planckian but small in string units,
reflecting that the resolution is a stringy effect. Hence these
black holes are called `small black holes', in contrast to the
`large' dyonic black holes, which already have a finite area
in the classical approximation.

Again it is crucial to deviate from the area law and to use
the generalized definition of black hole entropy (\ref{WaldSpec}), which 
results in  \cite{DKM}:
\be 
\Smacro = \frac{A}{4} + \mbox{Correction} = 
\frac{A}{4} + \frac{A}{4} = \frac{A}{2} =  4 \pi \sqrt{ \ft12  |q^2| }
= \Smicro + \cdots \;.
\ee
Thus we find that $\Smacro$ and $\Smicro$ are equal up to 
subleading contributions in the charges. 

We can try to improve on this result by including further 
subleading contributions to $\Smacro$. According to \cite{CdWM9906,CdWKM04}
the next relevant term is 
a non-holomorphic correction to the entropy which gives rise to
a term logarithmic in the charges:
\be
\Smacro =  4 \pi \sqrt{ \ft12  |q^2| } - 6 \log |q^2| \;.
\ee
This has the same form as $\Smicro$, but the coefficient of
the subleading term is different. This is, however, to be expected,
if $\Smacro$ and $\Smicro$ correspond to different ensembles.
The actual test consists of the following: take the
black hole free energy corresponding to the above $\Smacro$,
and evaluate the integral (\ref{MicroDeg}), or any candidate
modification thereof like (\ref{MicroDegOurs}), in a saddle point 
approximation and compare the result to $\Smicro$. 
This is, however not completely straightforward, because
the measure in (\ref{MicroDegOurs}) vanishes identically 
when neglecting non-holomorphic and non-perturbative corrections.
This reflects that the attractor points for electric 
black holes sit at the boundary of the classical moduli space
(the K\"ahler cone). This boundary disappears in the 
quantum theory, and non-holomorphic and 
non-perturbative corrections make the measure finite. But still,
the point around which one tries to expand does not correspond
to a classical limit. This might explain why there is still
disagreement for the term of the form $ \log |q^2|$ when
taking into account the leading non-holomorphic contribution
to the measure (\cite{CdWKM05}). Moreover, it appears
that \cite{Moore1,Moore2}, who take a different attitude
towards non-holomorphic contributions than \cite{CdWM9906,CdWKM05},
also find a mismatch for the logarithmic term. 
It is tempting to conclude that the conjecture 
(\ref{DiscLapl}) just
does not apply to small black holes. However, it was 
shown in \cite{Moore1,Moore2} that 
there is an infinite series of sub-subleading
contributions, involving inverse powers of the charges,
which matches perfectly! This suggest that there is a more 
refined version 
of the conjecture which applies to small black holes.
One possibility is that in order to have a simple 
relation between the macroscopic and mircoscopic entropy,
the latter has to be defined not with respect to the
microcanonical ensemble, but using another ensemble.
Concrete proposals are the `grand canonical ensemble'
used in \cite{Sen3,Sen4}
and the `redefined OSV ensemble' of \cite{ParTav}.

\section{Dyonic strings and `large' black holes}
\setcounter{equation}{0}

Let us finally briefly discuss $\ft14$ BPS black holes. 
While the leading order black hole entropy is (\ref{N=4tree}),
one can also derive a formula which takes into
account the non-perturbative and non-holomorphic 
corrections \cite{CdWM9906,CdWKM05}:
\be
\Smacro(q,p) = - \pi \left[
\frac{q^2 - i (S-\overline{S}) q \cdot p + |S|^2 p^2 }{
S + \overline{S}} - 2 \log [ (S + \overline{S})^6 |\eta(S)|^{24} ]
\right]_{\rm horizon}
\label{SmacroFull}
\ee
Here $S$ denotes the dilaton and $\eta(S)$ is the Dedekind
$\eta$-function. Recalling that the dilaton is related to the
string coupling $g_S$ by $S= \ft{1}{g_S^2} + i \theta$, and
using the expansion of the $\eta$-function,
\be
\eta(S) = - \ft1{12} \pi S - e^{- 2 \pi S} + {\mathcal O}(e^{-4\pi S})
\ee
we see that (\ref{SmacroFull}) includes an infinite series of
instanton corrections.

In order to show that (\ref{SmacroFull}) is invariant under T-duality
and S-duality, note that $q^2, p^2, p \cdot q$ and $S$ are invariant
under T-duality. Under S-duality $(q^2, p^2, p \cdot q)$ transforms
in the ${\bf 3}$ of $SL(2,\mathbbm{Z})$, while 
\be
S \rightarrow \frac{a S -i b}{icS + d} \;, \;\;\;
\left( \begin{array}{cc} 
a & b \\ c & d \\ 
\end{array} \right)  \in SL(2,\mathbbm{Z})\;.
\ee
It is straighforward to see that $(S+ \overline{S})^{-1}
(1, -i (S-\overline{S}), |S|^2)$ transforms in the ${\bf 3}$ of
$SL(2,\mathbbm{Z})$ so that
the first term of (\ref{SmacroFull})
is S-duality invariant. S-duality invariance of
the second term follows from the fact that $\eta^{24}(S)$ is a 
modular form of degree 12. Observe that the non-holomorphic 
term $\sim \log (S+\overline{S})$  is needed to make
the second term of (\ref{SmacroFull}) S-duality invariant.

The entropy formula (\ref{SmacroFull}) is not fully explicit, since
one cannot solve the attractor equations explicitly for 
the dilaton as a function of the charges. However, the dilaton
attractor equations take the suggestive form
\be
\frac{ \der \Smacro (q,p,S,\overline{S})}{\der S} = 0 \;.
\label{DilatonAttr}
\ee

Now we need to look for candidate microstates \cite{DVV}. Since 
these must carry electric and magnetic charge and must sit 
in $\ft14$ BPS multiplets, they cannot be fundamental string states.
However, the underlying ten-dimensional string theory contains
besides fundamental strings also solitonic five-branes, which
carry magnetic charge. Upon double dimensional reduction to six 
dimension these can become magnetic strings, which sit in 
$\ft12$ BPS multiplets. By forming bound states with fundamental strings
one can obtain dyonic strings forming $\ft14$ BPS multiplets. 
Further double dimensional reduction to four dimensions gives
dyonic zero-branes, which at finite coupling should correspond to
dyonic $\ft14$ BPS black holes.

Based on the conjecture that the world volume theory of the heterotic
five-brane is a six-dimensional string theory, one can 
derive a formula for the degeneracy of dyonic
states \cite{DVV}:
\be
d_{DVV}(q,p) = \oint_{C_3} d \Omega \frac{\exp i \pi ( Q, \Omega Q )}{
\Phi_{10}(\Omega)} \;,
\label{DVV}
\ee
where $\Omega$ is an element of the rank 2 Siegel upper half space
(i.e., a symmetric complex two-by-two matrix with positive definite
imaginary part). The vector

$Q = (q,p) \in \Gamma_{6,22} \oplus \Gamma_{6,22}$ combines the
electric and magnetic charges, $\Phi_{10}$ is the degree 2, weight 10
Siegel cusp form, and $C_3$ is a three-dimensional integration contour in
the Siegel upper half space. This formula is a natural generalization of
the degeneracy formula (\ref{ElectricStates})
for electric $\ft12$ BPS states. Recently, an alternative derivation
has been given \cite{SSY4}, which uses the known microscopic 
degeneracy of the five-dimensional D5-D1-brane bound state and
the relation between five-dimensional and four-dimensional black 
holes \cite{GHLS}.

It has been shown in \cite{DVV} that the saddle point value 
of the integral (\ref{DVV}) gives the leading order
black hole entropy (\ref{N=4tree}). More recently it
has been shown in \cite{CdWKM04} 
that a saddle point evaluation of $(\ref{DVV})$
yields precisely the full macroscopic entropy (\ref{SmacroFull}).
In particular, the conditions for a saddle point of the integrand
of (\ref{DVV}) are precisely the dilaton attractor equations 
(\ref{DilatonAttr}).

The natural next step is to investigate whether the 
microscopic state degeneracy (\ref{DVV}) is consistent 
with the symplectically covariant version (\ref{MicroDegOurs})
of (\ref{MicroDeg}). This can indeed be shown \cite{CdWKM05}
using the recent result of \cite{ShihYin}, who have evaluated
the mixed partition functions of BPS black holes in
$N=4$ and $N=8$ string compactifications. In particular,
the non-trivial measure found in \cite{ShihYin} can be obtained
from (\ref{MicroDegOurs}) by taking the limit of large charges
\cite{CdWKM05}.

\section{Discussion}
\setcounter{equation}{0}

We have seen that there is an impressive agreement between the
counting of supersymmetric string states and the entropy of supersymmetric
black holes. In particular, these comparisons are sensitive to the
distinction between the naive area law and Wald's generalized definition.
Moreover, there appears to be a direct link between the black hole 
entropy and the string partition function. This is a big leap forward
towards a conceptual understanding of black hole microstates. The stringy 
approach to black hole entropy 
also has its limitations. One is that in order to identify
the black hole microstates one has to extrapolate to asymptotically
small coupling. In particular, one does not really get a good understanding
of the black hole microstates `as such,' but only how
they look like in a different regime of the theory. A second, 
related problem is that one needs supersymmetry to have sufficent
control over the extrapolation, the state counting and the construction
of the corresponding black hole solutions. Therefore, quantitative 
agreement has only been established for supersymmetric black holes, 
which are charged, extremal black holes, and therefore
not quite relevant for astrophysics. It should be stressed, however,
that the proportionality between microscopic entropy and area
can be established by a variety of methods, including the 
string-black hole correspondence reviewed at the beginning. 
Moreover, the work of \cite{Sen:0506,Sen:0508} and \cite{KraLar}
suggests that the relation between black hole entropy 
and string theory states holds without supersymmetry.

The main limitation of string theory concerning quantum gravity
in general and black holes in particular is that its 
core formalism, string perturbation theory around on-shell
backgrounds, cannot be used directly to investigate the dynamics
of generic, curved, non-stationary space-times. 
But we have
seen that even within these limitations one can obtain 
remarkable results, which define benchmarks 
for other candidate theories of quantum gravity.
Historically, one important guideline for finding new physical 
theories has always been that one should be able to reproduce
established theories in a certain limit. This is reflected by
the prominent role that the renormalization group and 
effective field theories play in contemporary quantum field theory.
Concerning quantum gravity, it appears to be important 
to keep in touch with classical 
gravity in terms of a controlled semi-classical limit. Specifically,
a theory of quantum gravity should allow one to construct a 
low energy effective field theory, which contains the Einstein-Hilbert
term plus computable higher derivative corrections. Moreover,
if microscopic black hole states can be identified and counted, 
the question as to 
whether they obey the area law or Wald's generalized
formula should be answered.

The study of black holes has also lead to new ideas which will
hopefully improve our understanding of string theory. Notably we have seen
that there is a lot in common between supersymmetric black holes
and flux compactifications, in particular the role of 
variational principles. 
The relation between the black hole partition function and the
topological string partition function, and the interpretation of the latter
as a wave function shows that there is a kind of minisuperspace
approximation, which can be used to investigate the dynamics
of flux compactifications and quantum cosmology using a stringy
version of the Wheeler de Witt equation. This could not only 
improve the conceptual understanding of string theory, but would also
increase the overlap between string theory and canonical approaches
to quantum gravity.

More work needs to be done in order to further
develop the proposal made by \cite{OSV}. One key 
question is the relation 
between the macroscopic and microscopic entropies 
for small black holes, another one is the 
role of non-perturbative 
corrections in general \cite{Moore1,Moore2,ShihYin,CdWKM05}.
Future work will have to decide whether the relation 
discovered by \cite{OSV} is an exact or only an asymptotic
statement. Besides non-perturbative corrections also non-holomorphic
corrections are important.
We have discussed a concrete proposal for treating non-holomorphic
corrections based on \cite{CdWKM05}, but it is not obvious
how this relates in detail to the microscopic side, i.e., to the
topological string. 

\subsection*{Acknowledgment}
I would like to thank Bertfried Fauser and J\"urgen Tolkdorf for 
organising a very diverse and stimulating workshop. The original
results reviewed in this contribution were obtained in collaboration
with Gabriel Lopes Cardoso, Bernard de Wit and J\"urg K\"appeli,
who also gave valuable comments on the manuscript.


\begin{thebibliography}{1}
\bibitem{Sus} 
L.~Susskind, {\em Some Speculations about Black Hole Entropy in 
String Theory}. In C. Teitelboim (ed.), The Black Hole, 118.
hep-th/9309145.
\bibitem{HorPol} 
G.~Horowitz and J.~Polchinski, Phys. Rev. D 55 (1997) 6189, 
hep-th/9612146.
\bibitem{OSV}
H.~Ooguri, A.~Strominger and C.~Vafa, 
Phys. Rev. D 70 (2004) 106007, hep-th/0405146.
\bibitem{Moore1} 
A.~Dabholkar, F.~Denef, G.~W.~Moore, B.~Pioline, 
JHEP 08 (2005) 021, JHEP 08 (2005) 021, 
hep-th/0502157.
\bibitem{Moore2}
A.~Dabholkar, F.~Denef, G.~W.~Moore, B.~Pioline, 
hep-th/0507014. 
\bibitem{StrVaf}
A.~Strominger and C.~Vafa, Phys. Lett. B 379 (1996) 99, 
hep-th/9601029.
\bibitem{Gib}
G.~W.~Gibbons, {\em Supersymmetric Soliton States in Extended
Supergravity Theories}. In P.~Breitenlohner and H.~P~D\"urr 
(eds.), Unified Theories of Elementary Particles, Springer, 1982.
\bibitem{GibHul}
G.~W.~Gibbons and C.~M.~Hull, Phys. Lett. B 109 (1982) 190.
\bibitem{dWLPSvP}
B.~de~Wit, P.~G.~Lauwers, R.~Philippe, S.-Q. Su and
A.~Van~Proeyen, Phys. Lett. B 134 (1984) 37.
\bibitem{dWvP}
B. de Wit and A.~Van~Proeyen, Nucl. Phys. B 245 (1984) 89.
\bibitem{ACD} 
D.~V.~Alekseevsky, V.~Cortes and  C.~Devchand,
J. Geom. Phys. 42 (2002) 85, math.dg/9910091.
\bibitem{Kow-Glik}
J.~Kowalski-Glikman, Phys. Lett. B 150 (1985) 125.
\bibitem{CdWKM}
G.~L.~Cardoso, B.~de~Wit, J.~K\"appeli and T.~Mohaupt,
JHEP 0012 (2000) 019, hep-th/0009234.
\bibitem{FKS}
S.~Ferrara, R.~Kallosh and A.~Strominger, Phys. Rev. D 52 (1995)
5412, hep-th/9508072.
A.~Strominger, Phys. Lett. B 383 (1996) 39, hep-th/9602111.
S.~Ferrara and R. Kallosh, Phys. Rev. D 54 (1996) 1514,
hep-th/9602136.
S.~Ferrara and R. Kallosh, Phys. Rev. D 54 (1996) 1525,
hep-th/9603090.
\bibitem{CdWM}
G.~L.~Cardoso, B.~de~Wit and T.~Mohaupt,
Phys. Lett. B 451 (1999) 309, hep-th/9812082.
\bibitem{BCdWKLM}
K.~Behrndt, G.~L.~Cardoso, B.~de~Wit, R.~Kallosh, D.~L\"ust and T.~Mohaupt,
 Nucl. Phys. B 488 (1997) 236, hep-th/9610105.
\bibitem{Shm}
M.~Shmakova, Phys. Rev. D 56 (1977) 540, hep-th/9612076.
\bibitem{BLS}
K.~Behrndt, D.~L\"ust and W.~A.~Sabra,
Nucl. Phys. B 510 (1998) 264,hep-th/9705169. 
\bibitem{MSW}
J.~Maldacena, A.~Strominger and E.~Witten,
JHEP 12 (1997) 002, hep-th/9711053.
\bibitem{Vaf}
C.~Vafa, Adv. Theor. Math. Phys. 2 (1998) 207, hep-th/9711067.
\bibitem{deW1}
B.~de~Wit, Nucl. Phys. Proc. Suppl. 49 (1996) 191, hep-th/9602060.
\bibitem{TMhabil}
T.~Mohaupt, Fortsch. Phys. 49 (2001) 3, hep-th/0007195. 
\bibitem{BCOV}
M.~Bershadsky, S.~Cecotti, H.~Ooguri and C.~Vafa,
Comm. Math. Phys. 165 (1993) 311, hep-th/9309140.
\bibitem{AGNT}
I.~Antoniadis, E.~Gava, N.~S.~Narain and T.~R.~Taylor,
Nucl. Phys. B 413 (1994) 162, hep-th/9307158.
\bibitem{Torino}
T.~Mohaupt, Class. Quant. Grav. 17 (2000) 3429, hep-th/0004098.
\bibitem{Wald1}
R.~Wald, Phys. Rev. D 48 (1993) 3427, gr-qc/9307038.
\bibitem{Wald2}
V.~Iyer and R.~Wald, Phys. Rev. D 50 (1994) 846, gr-qc/9403028.
\bibitem{WittenBGI}
E. Witten, hep-th/9306122.
\bibitem{OVV}
H.~Ooguri, C.~Vafa and E.~Verlinde, hep-th/0502211.
\bibitem{GukSarVaf}
S.~Gukov, K.~Saraikin and C.~Vafa, hep-th/0509109.
\bibitem{CdWKM05}
G.~L.~Cardoso, B.~de~Wit, J.~K\"appeli and T.~Mohaupt, 
{\em in preparation}.
\bibitem{B1}
B.~de~Wit, hep-th/0503211.
\bibitem{B2}
B.~de~Wit, hep-th/0511261.
\bibitem{J}
J. K\"appeli, hep-th/0511221.
\bibitem{CdWM99}
G.~L.~Cardoso, B.~de~Wit and T.~Mohaupt, 
Fortsch. Phys. 48 (2000) 49, hep-th/9904005.
\bibitem{Sen:0506}
A.~Sen, hep-th/0506177.
\bibitem{Sen:0508}
A.~Sen, hep-th/0508042.
\bibitem{BCM}
K.~Behrndt, G.~L.~Cardoso and S.~Mahapatra, hep-th/0506251.
\bibitem{Str5d4d}
D.~Gaiotto, A.~Strominger and X.~Yin, hep-th/0503217,
hep-th/0504126.
\bibitem{KraLar}
P.~Kraus and F.~Larsen, JHEP 09 (2005) 034, 
hep-th/0506176.
\bibitem{HarMinMoo}
J.~A.~Harvey, R.~Minasian and G.~Moore, JHEP 09 (1998) 004.
hep-th/9808060.
\bibitem{GHLS}
M. Guica, L. Huang, W. Li and A. Strominger,
hep-th/0505188.
\bibitem{DGNV}
R.~Dijkgraaf, S.~Gukov, A.~Neitzke and C.~Vafa,
hept-th/0411073. 
\bibitem{Hitchin}
N.~Hitchin, math.dg/0107101, math.dg/0010054.
\bibitem{KapLou}
V.~Kaplunovsky and J.~Louis, Nucl. Phys. B 444 (1995) 191,
hep-th/9502077.
\bibitem{dWKLL}
B.~de~Wit, V.~Kaplunovsky, J.~Louis and D.~L\"ust,
Nucl. Phys. B 451 (1995) 53, hep-th/9504006.
\bibitem{CdWM9906}
G.~L.~Cardoso, B.~de~Wit and T.~Mohaupt, Nucl.Phys. B567 (2000) 87,
hep-th/9906094.
\bibitem{dWCLMR}
B.~de Wit, G.~L.~Cardoso, D.~L\"ust, T.~Mohaupt and S.-J.~Rey,
Nucl. Phys. B 481 (1996) 353, hep-th/9607184.
\bibitem{HarMor}
J.~A.~Harvey and G.~Moore, Phys. Rev. D 57 (1998) 2323,
hep-th/9610237.
\bibitem{Ver}
E. Verlinde, hep-th/0412139.
\bibitem{GS}
A.~A.~Gerasimov and S.~L.~Shatashvili, JHEP 0411 (2004) 074,
hep-th/0409238.
\bibitem{Dab}
A.~Dabholkar, Phys. Rev. Lett. 94 (2005) 241301,
hep-th/0409148.
\bibitem{Cvetic}
M.~Cvetic and D.~Youm, Phys. Rev. D 53 (1996) 584,
hep-th/9507090. M.~Cvetic and A.~Tseytlin, 
Phys. Rev. D53 (1996) 5619, Erratum-ibid. D55 (1997) 3907,
hep-th/9512031. 
\bibitem{SchwSen}
J.~Schwarz and A.~Sen, Phys. Lett. B 312 (1993) 105,
hep-th/9305185.
\bibitem{DLR}
M.~Duff, J.~T.~Liu and J.~Rahmfeld, Nucl.Phys. B 459 (1996) 125,
hep-th/9508094.
\bibitem{CCLMR}
G.~L.~Cardoso, G.~Curio, D.~L\"ust, T.~Mohaupt
and S.-J.~Rey, Nucl. Phys. B 464 (1996) 18, hep-th/9512129.
\bibitem{DH}
A.~Dabholkar and J.~Harvey, Phys. Rev. Lett. 63 (1989) 478.
A.~Dabholkar, G.~W.~Gibbons, J.~A.~Harvey and F.~Ruiz~Ruiz,
Nucl. Phys. B 340 (1990) 33.
\bibitem{HR}
G.~H.~Hardy and S.~Ramanujan, Proc. Lond. Math. Soc. 2 (1918) 75.
\bibitem{DKM}
A.~Dabholkar, R.~Kallosh and A.~Maloney,
JHEP 0412 (2004) 059
hep-th/0410076.
\bibitem{DVV}
R.~Dijkgraaf, E.~Verlinde and H.~Verlinde, Nucl.Phys. B 484 (1997) 543,
hep-th/9607026.
\bibitem{SSY4}
D.~Shih, A.~Strominger and X.~Yin, hep-th/0505094.
\bibitem{CdWKM04}
G.~L.~Cardoso, B.~de~Wit, J.~K\"appeli and T.~Mohaupt, 
JHEP 12 (2004) 075, hep-th/0412287.
\bibitem{Sen3}
A.~Sen, JHEP 05 (2005) 059, hep-th/0411255
\bibitem{Sen4}
A.~Sen, JHEP 07 (2005) 063, hep-th/0502126.
\bibitem{ParTav}
S.~Parvizi and A.~Tavanfar, hep-th/0508231.
\bibitem{ShihYin}
D.~Shih and X.~Yin, hep-th/0508174.
\bibitem{CdWKM0012}
G.~L.~Cardoso, B.~de~Wit, J.~K\"appeli and T.~Mohaupt,
Forschr. Phys. 46 (2001) 557, hep-th/0012232.



\end{thebibliography}
\end{document}